\documentclass[pdflatex,sn-mathphys-num]{sn-jnl}

\usepackage{graphicx}%
\usepackage{multirow}%
\usepackage{amsmath,amssymb,amsfonts}%
\usepackage{amsthm}%
\usepackage{mathrsfs}%
\usepackage[title]{appendix}%
\usepackage{xcolor}%
\usepackage{textcomp}%
\usepackage{manyfoot}%
\usepackage{booktabs}%
\usepackage{algorithm}%
\usepackage{algorithmicx}%
\usepackage{algpseudocode}%
\usepackage{listings}%
\usepackage{upgreek}

\theoremstyle{thmstyleone}%
\theoremstyle{thmstyletwo}%

\theoremstyle{thmstylethree}%

\begin{document}

\title[Article Title]{Terahertz Cavity Phonon Polaritons in Lead Telluride in the Deep-Strong Coupling Regime}


\author*[1,2,3]{\fnm{Andrey} \sur{Baydin}}\email{baydin@rice.edu}

\author*[4]{Manukumara Manjappa}\email{mmanjappa@iisc.ac.in}
\author[5,10]{Sobhan Subhra Mishra}
\author[6]{Hongjing Xu}
\author[7,8]{Jacques Doumani}
\author[7,8]{Fuyang Tay}
\author[7,8]{Dasom Kim}
\author[9]{Paulo H. O. Rappl}
\author[9]{Eduardo Abramof}
\author[5,10]{Ranjan Singh}
\author[12]{Felix G. G. Hernandez}
\author*[1,2,3,6,13]{Junichiro Kono}\email{kono@rice.edu}

\affil[1]{\orgdiv{Department of Electrical and Computer Engineering}, \orgname{Rice University}, \orgaddress{\state{Houston, TX 77005}, \country{USA}}}

\affil[2]{\orgdiv{Smalley--Curl Institute}, \orgname{Rice University}, \orgaddress{\state{Houston, TX 77005}, \country{USA}}}

\affil[3]{\orgdiv{Rice Advanced Materials Institute}, \orgname{Rice University}, \orgaddress{\state{Houston, TX 77005}, \country{USA}}}

\affil[4]{\orgdiv{Instrumentation and Applied Physics Department}, \orgname{Indian Institute of Science}, \orgaddress{\state{Bangalore, 560012}, \country{India}}}

\affil[5]{\orgdiv{Center for Disruptive Photonic Technologies}, \orgname{The Photonics Institute}, \orgaddress{\state{Nanyang Technological University, 639798}, \country{Singapore}}}

\affil[6]{\orgdiv{Department of Physics and Astronomy}, \orgname{Rice University}, \orgaddress{\state{Houston, TX 77005}, \country{USA}}}

\affil[7]{\orgdiv{Applied Physics Graduate Program, Smalley--Curl Institute}, \orgname{Rice University}, \orgaddress{\state{Houston, TX 77005}, \country{USA}}}

\affil[8]{\orgdiv{Department of Electrical and Computer Engineering}, \orgname{Rice University}, \orgaddress{\state{Houston, TX 77005}, \country{USA}}}

\affil[9]{\orgdiv{Group of Research and Development in Materials and Plasma}, \orgname{National Institute for Space Research}, \orgaddress{\state{São José dos Campos, São Paulo 12201-970}, \country{Brazil}}}

\affil[10]{\orgdiv{Division of Physics and Applied Physics, School of Physical and Mathematical Sciences}, \orgname{Nanyang Technological University, 637371}, \orgaddress{\country{Singapore}}}

\affil[12]{\orgdiv{Instituto de Física}, \orgname{Universidade de São Paulo}, \orgaddress{\state{São Paulo 05508-090}, \country{Brazil}}}

\affil[13]{\orgdiv{Department of Material Science and NanoEngineering}, \orgname{Rice University}, \orgaddress{\state{Houston, TX 77005}, \country{USA}}}


\abstract{Lead telluride is an important thermoelectric material due to its large Seebeck coefficient combined with its unusually low thermal conductivity that is related to the strong anharmonicity of phonons in this material. Here, we have studied the resonant and nonperturbative coupling of transverse optical phonons in lead telluride with cavity photons inside small-mode-volume metallic metasurface cavities that have photonic modes with terahertz frequencies. We observed a giant vacuum Rabi splitting on the order of the bare phonon and cavity frequencies. Through terahertz time-domain spectroscopy experiments, we systematically studied the vacuum Rabi splitting as a function of sample thickness, temperature, and cavity length. Under the strongest light--matter coupling conditions, the strength of coupling exceeded the bare phonon and cavity frequencies, putting the system into the deep-strong coupling regime. These results demonstrate that this uniquely tunable platform is promising for realizing and understanding predicted cavity-vacuum-induced ferroelectric instabilities and exploring applications of light--matter coupling in the ultrastrong and deep-strong coupling regimes in quantum technology.}


\maketitle

\section{Introduction}\label{sec1}
The exploration of light--matter interactions has significantly advanced in recent years, leading to the emergence of various innovative concepts and technologies. One of the most intriguing phenomena in this domain is strong coupling, wherein the exchange of energy between light (photons) and matter (excitons, phonons, or other excitations) occurs at a rate faster than their respective decay rates. This results in the formation of hybrid light--matter states known as polaritons. Much interest currently exists in searching for nonintuitive quantum states, phases, and phenomena in matter that are expected to arise as a result of nonperturbatively strong coupling with the cavity photons of the quantum vacuum field~\cite{Forn-DiazEtAl2019RMP,FriskKockumEtAl2019NRP,PeracaEtAl2020SaS,HubenerEtAl2021NM,SchlawinEtAl2022APR}. The so-called ultrastrong coupling (USC) regime is reached when the light--matter coupling strength, $g$, becomes a significant fraction of the bare resonance frequency of light and matter, $\omega_0$. Furthermore, when $g$ exceeds $\omega_0$, the light--matter hybrid enters the exotic regime of deep-strong coupling (DSC) with a large population of virtual photons~\cite{CiutiEtAl2005PRB}. In this regime, the light and matter can decouple, causing the breakdown of the Purcell effect~\cite{DeLiberato2014PRL}. For quantum information processing applications, protected qubits~\cite{NatafCiuti2011PRL}, ultrafast quantum gates~\cite{RomeroEtAl2012PRL}, and long-lasting quantum memories~\cite{StassiNori2018PRA} are predicted. In addition, nonlinear optical processes at low photon fluxes are expected to occur with unit efficiency~\cite{GarzianoEtAl2016PRL,KockumEtAl2017PRA}, and the matter ground state can be modified~\cite{LatiniEtAl2021PNASU,AppuglieseEtAl2022S,SchlawinEtAl2019PRL,SchulerEtAl2020SP,SchlawinEtAl2022APR,PeracaEtAl2020SaS,LuEtAl2024}. However, much of the extraordinary physics predicted to emerge in the DSC regime remains unobserved.

So far, the DSC regime has been experimentally realized in a few different solid-state platforms, including superconducting circuits~\cite{YoshiharaEtAl2017NP}, Landau polaritons in a two-dimensional electron gas~\cite{BayerEtAl2017NL,MornhinwegEtAl2024NC}, and plasmonic crystals~\cite{MuellerEtAl2020N}. Phonon--photon coupling presents another avenue for controlling phase transitions in solids~\cite{LatiniEtAl2021PNASU,LuEtAl2024} and modifying chemical reactions~\cite{HutchisonEtAl2012ACIE,ThomasEtAl2019S}. To date, only a handful of studies have achieved USC for photon- and plasmon-phonon coupling~\cite{Barra-BurilloEtAl2021NC,YooEtAl2021NP,ZhangEtAl2021PRR,LingEtAl2024NL,MekonenEtAl2024AP} due to the smaller dipole moments of phonons compared to electronic excitations. While various types of cavities have been employed to increase the strength of photon--phonon coupling, the DSC regime has remained elusive. 

Here, we present a novel experimental system for exploring DSC physics: terahertz (THz) cavity phonon polaritons in lead telluride (PbTe). PbTe has been known for its excellent thermoelectric properties~\cite{XiaEtAl2020PRX} and its incipient ferroelectricity~\cite{JiangEtAl2016NC}, both of which are associated with the strong interactions between electrons and lattice vibrations. We found that THz transverse optical (TO) phonons in PbTe films~\cite{BaydinEtAl2022PRL,KawahalaEtAl2023C} can strongly couple with THz photons in small-mode-volume metasurface (MS) cavities. We performed THz time-domain spectroscopy (TDS) measurements on several PbTe films with different thicknesses in MS cavities. At zero detuning, we observed a vacuum Rabi splitting ($2g$) that is greater than twice the cavity-phonon bare resonance frequency ($\omega_0$), signaling the entrance into the DSC regime. Specifically, the largest normalized coupling constant realized was $\eta \equiv g/\omega_0 = 1.25$ for a 1.6-$\upmu$m-thick film. The manifestation of such a high coupling strength promises this cavity--matter combination to be an excellent test bed for the studies of DSC science and opens up new opportunities to modify and control material properties using quantum vacuum fields.

\section{Results and Discussion}\label{sec2}
\begin{figure}
    \centering
    \includegraphics[width=\textwidth]{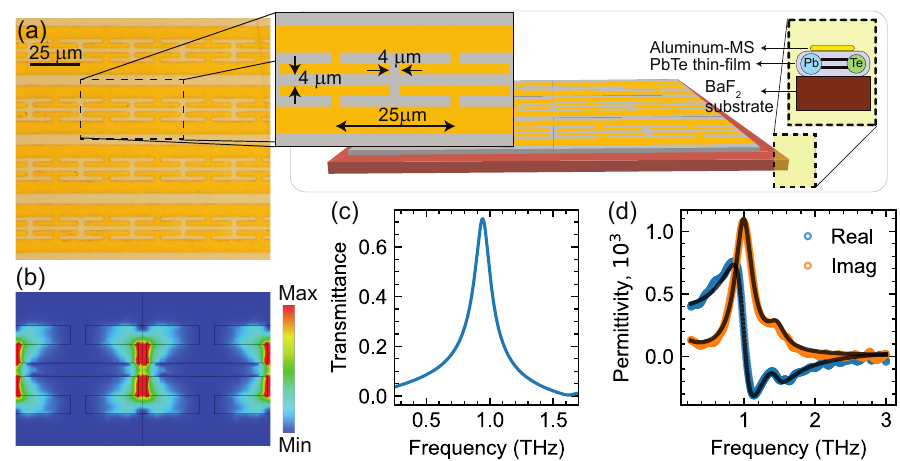}
    \caption{
    (a)~Microscope image and schematic of the sample where a PbTe thin film on a BaF$_{2}$ substrate is covered by a microstructured metal pattern representing a metasurface cavity with a resonance that can overlap with the transverse optical phonon of PbTe. The length of the cavity unit cell was 44\,$\upmu$m, 32\,$\upmu$m, 26\,$\upmu$m, and 14\,$\upmu$m for PbTe films with thicknesses of 100\,nm, 300\,nm, 500\,nm, and 1.6\,$\upmu$m, respectively.
    (b)~Numerical simulation of the electric field distribution at the resonance showing electric field enhancement in the 4-$\upmu$m gap of the metasurface unit-cell structure.
    (c)~A representative transmittance spectrum showing a cavity mode at about 1\,THz obtained from numerical simulations.
    (d)~Permittivity spectrum for the 1.6-$\upmu$m-thick PbTe film at room temperature. Open circles correspond to experimental data, and solid lines are fits to the data using the Drude-Lorentz model. 
    }
    \label{fig:schematic}
\end{figure}

The strength of resonant coupling between single-mode cavity photons and a two-level matter system can be written as $g = \mu_{12}E_\text{vac}/\hbar$, where $\mu_{12}$ is the dipole moment of the optical transition of the two-level system and $E_\text{vac}$ is the vacuum field strength in the cavity given by $E_\text{vac} = \sqrt{\hbar\omega_0 / 2\varepsilon_\text{r}\varepsilon_0 V_\text{m}}$. Here, $\varepsilon_\text{r}$ is the dielectric constant of the medium inside the cavity, $\varepsilon_0$ is the vacuum permittivity, and $V_\text{m}$ is the cavity mode volume. Thus, to maximize the coupling strength, we used PbTe, which has a large phonon Born effective charge~\cite{WaghmareEtAl2003PRB,OkamuraEtAl2022nQMa}, and a MS cavity with a small mode volume. Figure~\ref{fig:schematic}(a) illustrates the experimental configuration we employed, where we placed a metallic MS on a PbTe thin film on a BaF$_{2}$ substrate ($\varepsilon_\text{r}$ = 6.76). The gap dimension of the MS structure was 4\,$\upmu$m, providing strong confinement for the THz cavity photons that couple with the PbTe TO phonon mode. The MS was fabricated on PbTe using standard ultraviolet lithography techniques; see Methods for details about device fabrication.

An optical microscope image of the fabricated MS is shown in Figure~\ref{fig:schematic}(a). The basic unit cell of the MS with the cavity dimensions is schematically depicted in the inset. The MS design shows a single cavity (fundamental LC mode) resonance at 1\,THz for light polarized along its length. The LC mode of the MS cavity mode is designed in resonance with the phonon mode of PbTe at room temperature. The complementary MS design filters out the background transmission signal associated with the phonon resonance of the uncoupled region. In this cavity design, electromagnetic fields are confined and enhanced in the 4-$\upmu$m-gap regions [Figure~\ref{fig:schematic}(b)]. The shown electric field distribution was calculated using field monitors in the CST Microwave Studio Suite, as described in the Methods section. The vacuum electric field strength reaches 125\,V/m at the resonance frequency. Compared to conventional split-ring resonators, the two split gaps in the MS cavity provide better spatial overlap of the matter with the vacuum fields. We estimated the mode volume to be $10^{-4}$ ($\lambda/2)^{3}$ (or 348\,$\upmu$m$^{3}$) for resonance at 1\,THz. Figure~\ref{fig:schematic}(c) shows a resonance at 1\,THz of such an MS cavity with a $Q$-factor of 7. The resonance frequency of the cavity mode can be tuned by changing the length of the MS unit cell.

We used single-crystalline (111) PbTe films on BaF$_{2}$ substrates grown by molecular beam epitaxy with thicknesses ranging from 0.1\,$\upmu$m to 1.6\,$\upmu$m. The crystal group symmetry of PbTe is based on the rock-salt structure, where Pb and Te form interpenetrating face-centered cubic lattices with ionic bonds~\cite{XuEtAl2012NCa,BaydinEtAl2022PRL}. PbTe has a low-frequency TO phonon that softens with decreasing temperature and displays strong anharmonic lattice dynamics~\cite{RibeiroEtAl2018PRB,BaydinEtAl2022PRL}. To confirm the TO-phonon resonance and determine the complex optical constants, we used THz-TDS to characterize the films in free space. The measured permittivity of the 1.6-$\upmu$m-thick PbTe film at room temperature as a function of frequency is shown in Fig.~\ref{fig:schematic}(d). It was obtained by measuring the transmittance and applying the thin-film approximation~\cite{Tinkham1956PR,NussOrenstein}. The solid lines are the fitting using the Drude-Lorentz oscillator model~\cite{KawahalaEtAl2023C}. The prominent peak observed at around 1\,THz corresponds to the TO$_1$ phonon, which we use to create polariton states in cavities. The small feature at 1.4\,THz corresponds to the TO$_2$ phonon mode, which arises from the strong anharmonic interaction between the longitudinal acoustic (LA) and TO phonons~\cite{DelaireEtAl2011NM}.

\begin{figure}[ht!]
    \centering
    \includegraphics[width=\textwidth]{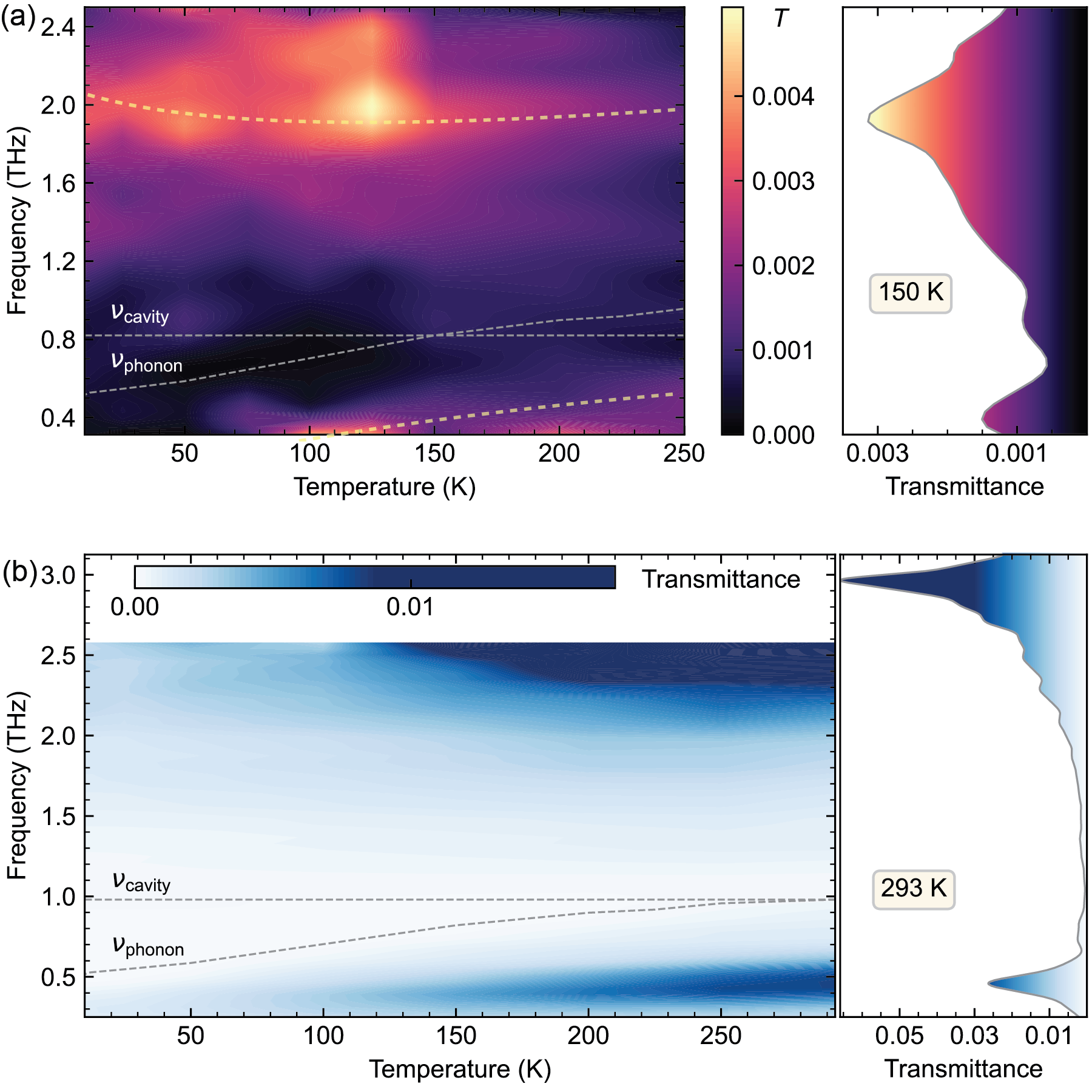}
    \caption{
    (a)~Measured transmittance spectra for the PbTe-MS structure with a PbTe film thickness of 300\,nm at various temperatures.  The MS cavity mode frequency was designed to equal the TO phonon frequency at 150\,K. The yellow dashed lines (in the left subfigure) are guides to the eye, indicating the lower and upper polariton branches. The right part of the figure shows the transmittance spectrum at 150\,K, corresponding to zero detuning. 
    (b)~Measured transmittance spectra for the PbTe-MS structure with a PbTe film thickness of 1.6\,$\upmu$m at various temperatures. The MS cavity mode frequency was at 1\,THz at all temperatures, so the detuning was zero at room temperature and increased with decreasing temperature. The left part shows a temperature-dependent transmittance color map and the right part shows the spectrum at room temperature, showing a giant Rabi splitting indicating DSC.
    }
    \label{fig:temperature}
\end{figure}

Using maskless photolithography followed by aluminum deposition with e-beam evaporation, we fabricated MS cavities on the PbTe film samples and characterized them by THz-TDS; see Methods for details. Temperature-dependence transmittance spectra for the PbTe films with different thicknesses are shown in Figure~\ref{fig:temperature}. At lower temperatures, the TO phonon mode softens~\cite{BaydinEtAl2022PRL} as indicated by the dashed line in Figure~\ref{fig:temperature}[Left]. 
Temperature-dependent polaritonic anticrossing was observed for the sample with a PbTe thickness of 0.3\,$\upmu$m. Figure~\ref{fig:temperature}(a) shows a color map consisting of experimental transmittance spectra at various temperatures. The MS cavity was designed such that zero detuning occurs at 150\,K. The bright regions in the color map indicate the lower and upper polariton branches, which are also shown by yellow dashed lines. 
The right part of Figure~\ref{fig:temperature})(a) shows a slice of the color map at 150\,K. The vacuum Rabi splitting is $\sim$1.5\,THz, which yields a normalized coupling strength $\eta$ = 0.91, approaching the DSC regime of the light--matter coupling. 

For the PbTe film with a thickness of 1.6\,$\upmu$m, we observed a clear anticrossing behavior with $\eta$ reaching 1.25 at room temperature, as shown in Figure~\ref{fig:temperature}(b). At 293\,K, the cavity mode and the TO phonon mode were in resonance at 1\,THz. 
Softening of the phonon mode to a shift in the polaritonic branches, with the lower polariton mode going below 0.5\,THz and the upper polariton at 3\,THz with decreasing temperature.  

\begin{figure}[h]
    \centering
    \includegraphics[width=\textwidth]{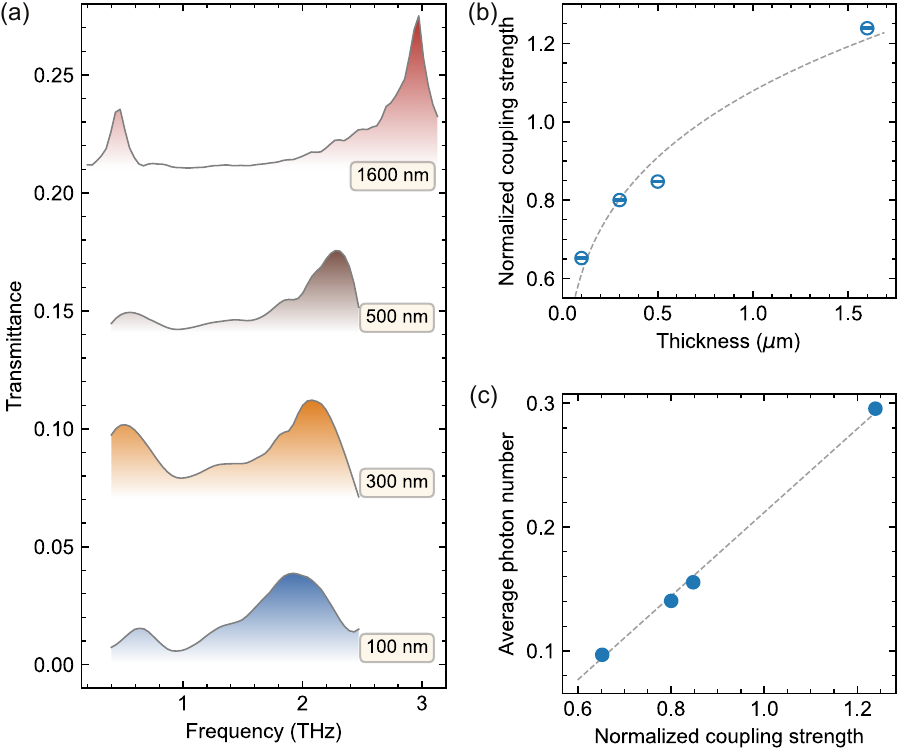}
    \caption{
    (a)~Transmittance spectra for PbTe films of varying thicknesses coupled with the MS designed at 1\,THz. The vacuum Rabi splitting between the two polaritonic peaks increases with the thickness of the PbTe film. 
    (b)~The normalized coupling strength ($\eta$) as a function of PbTe thickness, obtained from experimental data in (a) and shown by circles. The dashed line is a guide to the eye. 
    (c)~The average photon number in the polaritonic ground state corresponding to the measured coupling strength $\eta$ shown in (b).
    }
    \label{fig:splitting}
\end{figure}

In addition, we studied the PbTe film thickness dependence of the light--matter coupling and the corresponding $\eta$ at room temperature with the same MS structure. Figure~\ref{fig:splitting}(a) shows transmittance spectra for the PbTe-MS structure for different PbTe film thicknesses from 0.1\,$\upmu$m to 1.6\,$\upmu$m. The frequency of the uncoupled cavity resonance and the TO phonon mode of PbTe was 1\,THz. The spectra show two pronounced peaks corresponding to the lower and upper polariton branches, respectively. As the film thickness increases, the splitting between these two peaks -- the Rabi splitting -- increases. It varies from about 1.4\,THz to 2.5\,THz, which translates into the normalized coupling strength, $\eta$, changing from 0.65 (the USC regime) to 1.25 (the DSC regime). The dependence of the normalized coupling strength on the PbTe film thickness is summarized in Figure~\ref{fig:splitting}(b) at room temperature. The blue circles represent the results of experimental measurements, while the dashed line is a guide to the eye. The normalized coupling strength does not exactly follow the square root dependence on the film thickness as one would expect from the Dicke cooperativity enhancement factor of $\sqrt{N}$, where $N$ is the number of oscillators~\cite{YahiaouiEtAl2022NL}. The reason for this discrepancy is the fact that these films of PbTe grown on BaF$_2$ exhibit thickness-dependent optical properties due to unintentional doping during film growth~\cite{KawahalaEtAl2023C}. Nevertheless, these results are particularly significant, as they demonstrate that the system can be tuned between the USC and DSC regimes by simply varying the film thickness. Even for the smallest thickness of 100\,nm, the system is in the USC regime, and the DSC regime is achieved for the thickest measured sample of 1.6\,$\upmu$m. As mentioned earlier, the coupling strength is also observed to be temperature-dependent due to the softening of the TO phonon mode in this material. As shown in Figure~\ref{fig:temperature}(b) and Figure~\ref{fig:splitting}(b), the normalized coupling strength, $\eta$, is 0.91 at 150\,K and 0.8 at 293\,K. We also performed calculations for the same MS design on a 1.6\,$\upmu$m-thick PbTe film at 12\,K (not shown) and obtained $\eta = 2.2$.

In the DSC regime with $\eta$ = 1.25, the light--matter coupled system is expected to have a significant virtual photon population in the ground state, $|G\rangle$. We model the system by the Hopfield Hamiltonian, taking into account the antiresonant interactions and the diamagnetic terms, i.e., $\hat{H}_\text{anti} = \hbar \sum_{k} [\emph{i}g_{k} (\hat{a}_{k}\hat{b}_{-k} - \hat{a}_{k}^{+} \hat{b}_{-k}^{+}) + D_{k} (\hat{a}_{k}\hat{a}_{-k} + \hat{a}_{k}^{+} \hat{a}_{-k}^{+})]$, where $\hat{a}_{k}$ and $\hat{b}_{k}$ represent the annihilation operators for the photons and matter for the $k$-th excitation. For a parabolic confinement potential with the diamagnetic term $D_{k} \simeq 1.04 g_{k}^{2}/\omega_{12}$ and $\omega_{\text{cav},k} = \omega_{12}$, we obtain the ground state virtual photon population of $\langle G|\hat{a}_{k}^{+}\hat{a}_{k}|G\rangle = 0.3$ for $\eta = 1.25$.  Figure~\ref{fig:splitting}(c) presents the average photon number ($\langle G|\hat{a}_{k}^{+}\hat{a}_{k}|G\rangle$) in the polaritonic ground state as a function of normalized coupling strength, $\eta$. This is a key parameter in DSC systems, as the presence of a large number of virtual photons in the ground state leads to nontrivial quantum effects, such as the modification of the materials' ground state and the breakdown of the Purcell effect. In this study, the photon population increases from $\langle G|\hat{a}_{k}^{+}\hat{a}_{k}|G \rangle$ = 0.09 for the coupling strength $\eta$ = 0.65 and reaches a significant value of 0.3 for $\eta$ = 1.25, further confirming that the system operates in the DSC regime. The observation of DSC in the PbTe--MS cavity system with a finite photon population in the ground state suggests that the system has the potential to control phase transitions through cavity--photon interactions, opening up new possibilities for developing quantum materials with engineered quantum phases and properties.


\section{Conclusion}
In conclusion, we have experimentally investigated the coupling between the cavity mode of a small-mode-volume MS and the TO phonon mode in thin films of PbTe using THz time-domain spectroscopy. In transmittance spectra, we observed lower and upper polaritons with a vacuum Rabi splitting that is on the order of twice the bare cavity mode and phonon resonance frequencies for film thicknesses of 0.1–1.6\,$\upmu$m. Thus, by varying the thickness of the PbTe film, we were able to obtain values of the normalized coupling constant $\eta$ ranging from 0.6 to 1.25 that correspond to the USC and DSC regimes of light--matter interaction. These results indicate that the phonon polaritonic platform using MS plasmonic cavities is promising for realizing and understanding predicted exotic cavity-vacuum-induced effects, such as ferroelectric phase transitions, as well as for exploring applications of USC/DSC phenomena in quantum technology. Furthermore, the approach we described here for THz cavity-phonon polaritons can be applied to a wide range of materials with large phonon Born effective charges, such as ferroelectrics, piezoelectrics, and complex oxides, which exhibit strong coupling between lattice vibrations and electric fields. 

\section*{Methods}
\subsection*{PbTe films} Single-crystalline (111) PbTe films were grown by molecular beam epitaxy (RIBER 32P MBE system) on BaF$_2$ substrates. The (111)-oriented BaF$2$ substrate was acquired from Korth Krystalle GmbH. A total of four samples have been used with thicknesses of 0.1, 0.3, 0.5, and 1.6 $\mu$m. More details on the MBE growth of PbTe on BaF$2$ can be found in the literature~\cite{KawahalaEtAl2023C,RapplEtAl1998JoCG}. 

\subsection*{Metasurface fabrication}
The metasurfaces were fabricated using the photolithography technique in the maskless photolithography system Bruker SF-100 Lightning. We used AZ nLOF 2020 photoresist and AZ 300MIF developer. We deposited 150\,nm Aluminum and carried out the lift-off procedure using acetone. 

\subsection*{Terahertz measurements} Over the course of the project, the samples were measured with home-built and commercial (TOPTICA) THz time-domain spectrometers. The home-build THz setup provided a broadband THz pulse, containing frequencies between 0.25\,THz and 2.5\,THz, generated by a photoconductive antenna (PCA) and detected via electro-optic sampling in a ZnTe crystal using 800\,nm pulses from the output of a mode-locked Ti:sapphire laser with a repetition rate of 80\,MHz. The TOPTICA setup is PCA-based and provides bandwidth of up to 4\,THz. All measurements were carried out in transmission geometry. 

\subsection*{Electromagnetic simulations} The unit cell of the metasurface (MS) was designed and simulated in the frequency domain using the CST microwave studio suite. Before studying the coupling between cavity photons and phonons, the uncoupled resonance of the MS cavity with an MS-on-BaF$_2$ sample design was numerically characterized. For simulations, the aluminum layer (Al) of the thickness of 200\,nm was modeled as a lossy metal with the conductivity of $3.56\times10^{7}$ S/m. A calculated transmittance spectrum exhibiting the uncoupled resonance is shown in Fig.~\ref{fig:schematic}(c). The frequency of the cavity mode can be tuned to match the PbTe phonon mode by changing the length of the MS unit cell. The terahertz conductivity of PbTe thin films of varying thicknesses across various temperatures was obtained using terahertz time-domain spectroscopy and was used to simulate the cavity photon-matter phonon coupling to show the splitting.

\backmatter


\bmhead{Acknowledgements}
Authors acknowledge support from the U.S. Army Research Office (through Award No. W911NF2110157), the Gordon and Betty Moore Foundation (through Grant No. 11520), and the Robert A. Welch Foundation (through Grant No. C-1509). M.M. acknowledges support from IE/CARE-23-0303 and SERB (SP/SERB-23-0286) grants. F.G.G.H. acknowledges financial support from the Brasil@Rice Collaborative Grant and the São Paulo Research Foundation (FAPESP) grant nos. 2021/12470-8 and 2023/04245-0. This work was done in part using resources of the Shared Equipment Authority at Rice University.

\bibliography{2-references}

\end{document}